*Magnetic Nanoparticles as Label-Free Dual-Function Nanoheaters and Nanothermometers*

*Alejandro Venegas-Gomez, Pablo Palacios-Alonso, Cristina C. Carrizo, Julieta Velasco Martínez-Pardo, Olmo Gómez-Rubio, Sedef Ozel-Okcu, Rafael Delgado-Buscalioni , Sebastian A. Thompson\*, and Francisco J. Terán\**

*A.Venegas-Gómez, P. Palacios-Alonso, C.Carrizo, J. Velasco Martínez-Pardo, S.Thompson, F.J.Terán*

IMDEA Nanociencia. Campus Universitario Cantoblanco, 28049. Madrid, Spain

E-mail: Sebastian.thompson@imdea.org

E-mail: francisco.teran@imdea.org

*P. Palacios-Alonso, R. Delgado-Buscalioni*

Dpto. Física Teórica de la Materia Condensada, Universidad Autónoma de Madrid, Madrid, Spain

*S.Thompson, F.J.Terán*

Nanobiotecnología (IMDEA Nanociencia), Unidad Asociada al Centro Nacional de Biotecnología (CSIC). 28049 Madrid,Spain

*F.J.Terán*

Nanomateriales Avanzados (IMDEA Nanociencia), Unidad Asociada al Instituto de Ciencia de Materiales de Madrid-CSIC, 28049 Madrid, Spain

*R. Delgado-Buscalioni*

Condensed Matter Institute, IFIMAC, Madrid, Spain

*Olmo Gómez-Rubio*

Departamento de Física Matemática y de Fluidos, UNED, 28232 Las Rozas, Madrid

*Olmo Gómez-Rubio, Sedef Ozel-Okcu*

Nanotech Solutions, Carretera Madrid 23, 40150 Villacastín, Spain

Funding: This work was partially funded by Europe Horizon (STRIKE, GA num. 101072462), Spanish Ministry of Science and Innovation (CNS2024-154353., PID2024-158994OB-C41, PID2024-158994OB-C44, PID2020-117080RB-C51, PID2020-117080RB-C53 ) and Comunidad de Madrid (MAG4TIC, TEC-2024/TEC-380). IMDEA Nanociencia acknowledges the "Severo Ochoa" program for Centres of Excellence in R&D (CEX2020-001039-S). AVG and PPA thank Agencia Estatal de Investigación for FPI fellowships CEX2020-001039-S-21-4 PRE2021-100202. and P. Palacios Alonso: CEX2020-001039-S-21-4 PRE2021-100485, respectively.

Keywords: nanothermometers, nanoheaters, magnetic nanoparticles, dynamical magnetization, photothermal actuation,

**Abstract**

Heat generation and temperature reading at the nanoscale have attracted increasing attention due to their direct relevance in thermal therapeutic approaches. Consequently, huge progress has been made toward the design of dual-function nanoplatforms that integrate heating and thermometry capabilities at the nanoscale. However, in most cases, dual nanoheater–nanothermometer platforms rely either on specifically engineered materials or on complex readout schemes, which limits translational potential due to complex implementation procedures. To overcome these challenges, we present a methodology for directly extracting temperature information based on dynamical magnetization measurements of cobalt ferrite magnetic nanoflowers. We demonstrate that these nanocrystals monitor temperature changes through variations in their magnetization cycles measured under alternating magnetic fields. Importantly, this thermometric functionality is preserved after surface functionalization and under chemical variations in the nanoparticle environment. Interestingly, we show that we can simultaneously generate heat and report temperature changes within the same agent. This is thanks to the photothermal conversion of cobalt ferrite nanoparticles subjected to near infrared irradiation and the tight reported relationship between magnetization dynamics and Brownian relaxation. Together, these results establish cobalt ferrite magnetic nanoparticles as a label-free platform for simultaneous heat generation and intrinsic temperature readout, enabling real-time nanoscale thermal control.

**Introduction**

*Nanothermometers* enable quantification of the temperature of the object of study at the nanoscale, a regime inaccessible to traditional thermometry, which relies on macroscopic measurements [1]. Exploration of non-invasive nanoscale thermal reading methodologies is crucial for understanding and guiding the development of new thermal therapies based on nanomaterials in biomedicine [2–4]. For this purpose, most of the methodologies to probe thermal phenomena at the nanoscale rely on optical techniques, such as fluorescence-based methods, fluorescence polarization anisotropy, fluorescence lifetime measurements, and related techniques [1,5–11]. In addition, magnetic nanoparticles have also been introduced as nanothermometers, most implementations require specialized magnetic instrumentation for harmonic or phase-sensitive detection, which limits experimental accessibility and integration with standard optical and biological imaging platforms[12–14]. In parallel, *nanoheaters* can convert various forms of external energy into localized heat, enabling controlled thermal perturbations with high spatial precision [15]. Such localized heating released by nanomaterials has attracted considerable interest in biomedicine, particularly for applications such as magnetic hyperthermia and/or photothermal therapies allowing to remove malignant cells and tissues [16]. A broad range of nanoheating strategies have been explored, including plasmonic nanoparticles activated by light, carbon-based nanomaterials, molecular photothermal agents and magnetic nanomaterials [17,18]. In this context, magnetic nanoheating has emerged as a powerful alternative led by magnetic nanoparticles. The latest two different external stimuli in order to generate heat: external alternating magnetic fields or light irradiation [19]. These approaches offer deep tissue penetration, remote actuation, and excellent temporal control. Despite the rapid and separate development of *nanoheaters* and *nanothermometers*, integrating both functionalities within a single platform remains challenging. Most dual nanoheating–thermometry strategies reported rely on specifically engineered compounds or complex architectures that require dedicated chemical synthesis, or on readout schemes that are not easily implemented, such as magnetic-field–driven rotational dynamics or anti-Stokes spectral analysis [12,20–23].

To overcome these limitations, here we present a thermal reading methodology based on the dynamical magnetization measurements of the cobalt ferrite nanoflowers (MNPs) [24]. In this way, we enable temperature readout from the same material responsible for heat generation. When Brownian relaxation prevails, any phenomena influencing MNP diffusion in liquid media will be reflected on AC magnetic hysteresis loops recorded under alternating magnetic fields. In particular, the thermal dependence of water viscosity enables temperature readout from MNPs. We demonstrate that temperature can be directly extracted in simultaneous to local heating generation. Hence, our evidences underline the dual capabilities of MNPs acting as direct thermometer and heat with no need for additional thermal sensing labels. Thus, we propose to exploit AC magnetic hysteresis loops to transduce thermal information from MNPs whose relaxation mechanisms are dominated by Brownian process. The variations of the AC magnetic hysteresis area accurately reflect macroscopic temperature.

**Result and discussion**

In order to report how the temperature information can be label-free extracted from the dynamical magnetization of MNPs subjected to alternating (AC) magnetic fields ranging from 10 till 100 kHz and up to 24 kA/m. For this purpose, three commercial MNP formulations were employed: plain MNPs, dextran coated MNPs (D-MNPs) and BSA coated D-MNPs (BSA-D-MNP) described in Experimental Section. Recent theoretical works[25] allow us to determine how the dynamic magnetization is influenced by the temperature using numerical calculations that can predict the temperature dependence of the experimental AC hysteresis loops (Fig. 1a) for plain MNPs [25]. This theoretical approach is based on an accurate phenomenological equation (<1% error) derived from exact Focker-Planck results and validated against experiments. This relation describes the frequency dependence of the AC magnetization loop area, assuming that Brownian relaxation is the only active mechanism (see Ref. [25] for more information). This assumption holds for particles with high magnetic anisotropy, such as cobalt ferrites [26]. The method includes an automated fitting procedure that allows the simultaneous estimation of the magnetic moment of the particles and their Brownian relaxation time.

$$\tau_B = \frac{4\pi\eta r_h^3}{k_B T}$$

Here, $r_h$ is the hydrodynamic radius of the particles, $k_B$ is the Boltzmann constant, T is the temperature of the medium, and η(T) is the temperature-dependent viscosity of the solvent. Once the Brownian relaxation time is determined from the fitting procedure, any of the associated parameters can be inferred provided the others are known. More details about this theoretical model can be found in Supporting Information 2. The results of the previous theoretical prediction are shown in the following figure, along with experimental data.

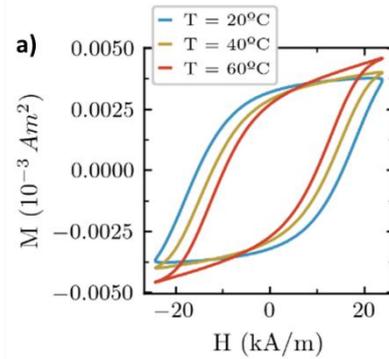

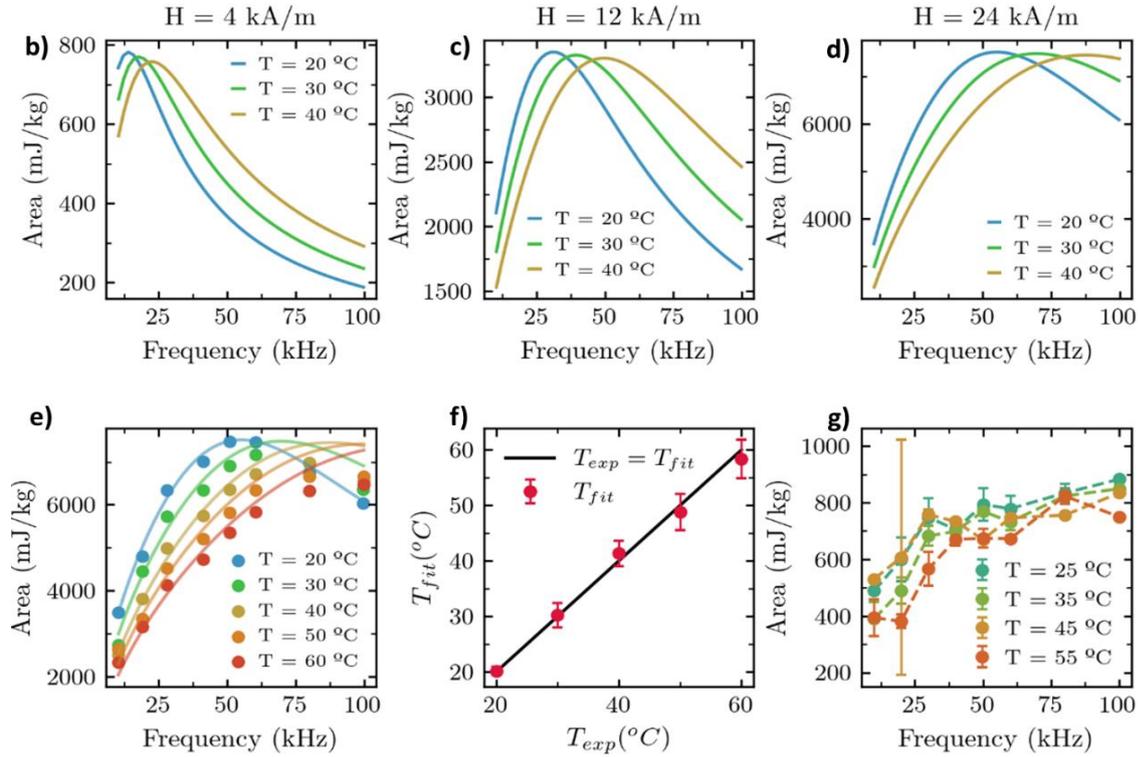

**Figure 1. Frequency dependence of AC magnetic hysteresis area at different field intensities (4, 12 and 24 kA/m) and temperatures (from 20 to 60ºC) for MNPs:** (a) AC magnetic hysteresis loops of MNPs at 20°C (blue), 40°C (green) and 60°C (red). (b–d) Theoretically predicted AC magnetic hysteresis area (A) as a function of different field frequencies for plain MNPs computed at different temperatures (20, 30, and 40°C) under increasing applied magnetic field 4 kA/m, 12 kA/m and 24 kA/m (e) Experimentally measured (dots) AC magnetic hysteresis area as a function of field frequency at different temperatures (20–60°C), compared with numerical predictions (solid lines) computed using the best-fitting temperature (T_fit) for the MNPs (f) Comparison between the experimental temperature at which the measurements were performed (T_exp) and the fitted temperature using the theoretical model that best reproduces the experimental data (T_fit). (g) Frequency-dependence of the AC magnetic hysteresis area for Synomag® MNPs (energy dissipation is dominated by Néel relaxation) at 24 kA/m and different temperatures. MNPs are dispersed in DDW at 1 g/L. AC magnetic field frequency values range from 10 up to 100 kHz. The particle parameters ($r_h$ and $m$) employed in the numerical calculations are given in Supporting Information 2.

As shown in Figure 1b-d, the magnetic energy absorption area depends on both the field frequency and the applied field intensity. At lower field intensities (4 kA/m), a well-defined maximum is observed at relatively low frequencies (between 20-30 kHz), which shifts toward higher frequencies (> 40 kHz) on increasing field intensity to 24 kA/m. Also, temperature progressively shifts the maximum to higher frequencies in all cases. This behavior with temperature values at 20ºC are consistent with a relaxation mechanism dominated by Brownian rotation of the magnetic nanoparticles in water [27]. The strong dependence on field amplitude and the broadening of the absorption spectra further supports a Brownian contribution, while a minor Néel relaxation contribution cannot be excluded [25]. To explore this behavior experimentally, as shown in Fig. 1e-f, we extended the temperature range to 60°C and compared it using MNPs in their plain form (MNPs). There is a precise match between experimental and theoretical (Figure 1f). In contrast, Synomag® iron oxide nanoparticles, for which energy dissipation is dominated by Néel relaxation [26], show different behavior under similar experimental conditions. Magnetite nanoflowers show a linear increase of AC magnetic area on increasing field frequency for all temperatures (Figure 1g). The absence of a well-defined temperature-dependent maximum highlights a fundamentally different relaxation mechanism from that in Brownian-dominated systems and supports the mechanistic interpretation of the Néel-dominated systems (see Figure SI1).

We then conducted a detailed analysis of the nanothermometric properties as a function of field frequency to identify the range over which MNPs exhibit a linear relationship with temperature. The corresponding results are shown in Figure 2a–c (To maximize the signal-to-noise ratio, all subsequent experiments were performed at an applied magnetic field amplitude of 24 kA/m).

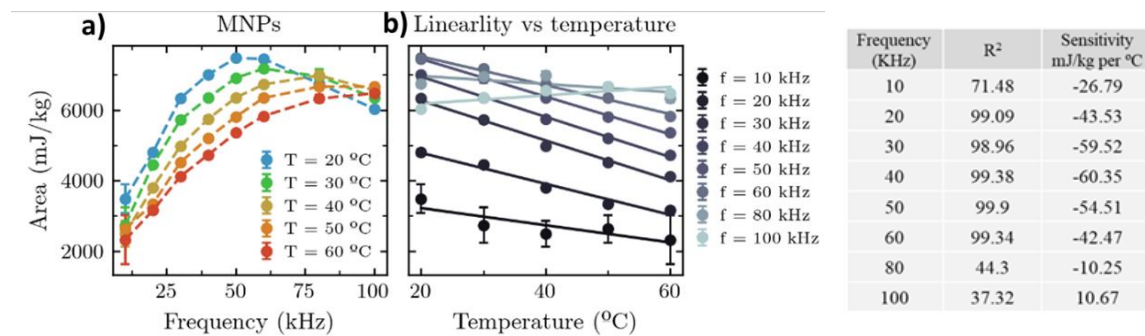

**Figure 2. Frequency dependence of AC magnetic hysteresis area at different temperatures (from 20 to 60°C)-for MNPs (a)** AC magnetic hysteresis area as a function of field frequency measured at 24 kA/m and different temperatures ranging from 20 till 60°C. **(b)** AC magnetic hysteresis area as a function of temperature for selected field frequencies ranging from 10–100 kHz extracted from (a). Table 1: Table showing the coefficient of determination ($R^2$) and temperature sensitivity as a function of field frequency.

As shown in Figure 2a, the effective magnetic response typically decreases with increasing temperature due to enhanced Brownian relaxation, which accelerates random rotational motion and reduces net alignment of magnetic moments. However, only a limited range of field frequencies (20–60 kHz) holds a linear relationship between temperature and magnetic hysteresis area (Figure 2b-c), with a maximum temperature sensitivity (the derivative of area respect temperature) observed between 30 and 50 kHz (table 1). These results clearly demonstrate that the local temperature of the nanoparticle environment can be precisely determined by measuring the AC magnetic hysteresis area of MNPs.

After assessing MNP's performance as nanothermometers in water, we investigated how the nanoparticle surface may influences their diffusion and consequently, the MNP dynamical magnetization. To this end, we explored two complementary scenarios: (1) we examined how MNP surface influences MNP diffusion and consequently the MNP dynamical magnetization;(2) we assessed how physico-chemical parameters of the MNP medium, such as viscosity or protein concentration influence the MNP dynamical magnetization.

Figure 3 shows the frequency dependence of AC magnetic hysteresis area at 24 kA/m at different temperatures ranging from 20 to 60ºC for the different surface: plain MNPs, dextran coated MNPs (D- MNPs) and bovine serum albumin coated MNPs (BSA-D-MNPs).

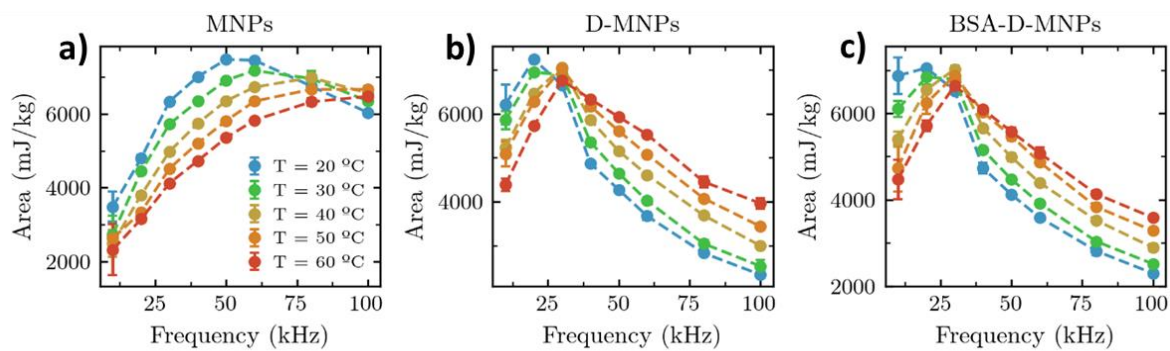

**Figure 3. Frequency dependence of AC magnetic hysteresis area at different temperatures (from 20 to 60ºC)-for distinct MNPs surfaces.** AC magnetic hysteresis-loop area as a function of field frequency measured at 24 kA/m and different temperatures (20–60°C) for (a) MNPs, (b) D–MNPs, and (c) BSA–D–MNPs.

As observed, the AC magnetic area trends present a similar thermal behavior between the three MNPs, but the maximum peak-value shifts depending on MNP surface. While the peak in area (maximum) of plain MNPs is located at frequencies around 50-60 kHz (at 20ºC), D- MNPs peak at 20 kHz and BSA-D-MNPs at 10-20 kHz. Consequently, their thermal response is differently expressed. While plain MNPs depict a strong thermal dependence with the field frequency range between 10 and 50 kHz, D-MNPs and BSA-D-MNPs presents largest variations above 40 kHz. This behavior is related to the different MNP diffusion depending on the surface coating (MNPs hydrodynamic diameter of 38.5 ± 0.1 with a low PDI of 0.04 ± 0.01 while D-MNPs hydrodynamic size increased to 56.7 ± 0.1 accompanied by a modest increase in PDI to 0.07 ± 0.01). Indeed, the observed shift of the maximum-area toward lower frequencies is due to the substantially larger hydrodynamic diameter in presence of Dextran with respect to plain MNPs, which reduces the MNP rotational Brownian diffusion. In this sense, AC magnetic hysteresis loops clearly discern the magnetic relaxation dynamics under different thermal environments and for different particle sizes[28]. Subsequent conjugation with BSA induces only minor additional changes, indicating that the dextran coating already imposes the dominant alteration of local rotational freedom.

We then performed the same analysis for D-MNPs, and the results are shown in the following figure (The analysis for BSA–D-MNPs is shown in Figure SI2).

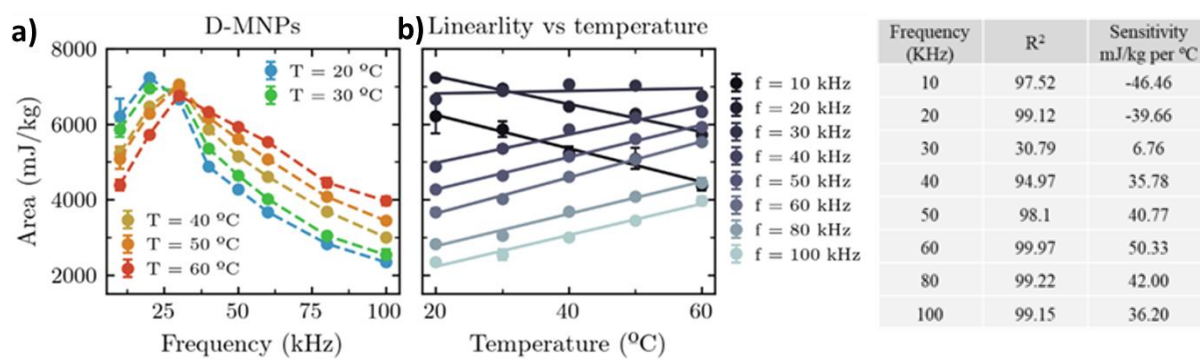

**Figure 4. Frequency dependence of AC magnetic hysteresis area at different temperatures (from 20 to 60ºC)-for D-MNPs.**
(**a**) AC magnetic hysteresis area as a function of field frequency measured at 24 kA/m and different temperatures (20–60°C). (**b**) Magnetic hysteresis area as a function of temperature for selected field frequencies (10–100 kHz), illustrating the frequency-dependent sign and magnitude of the thermometric response. Table 2: Table showing the coefficient of determination ($R^2$) and temperature sensitivity as a function of field frequency.

As shown in Fig. 4a the magnetic response reaches its maximum around 30 kHz, a value which does not largely vary with the temperature. Notably, quasi-linear scaling of the hysteresis area with temperature are recovered below and above this peak frequency, respectively presenting positive and negative slopes with the temperature. This "bidirectional" temperature dependence is to our knowledge, unprecedented in magnetic nanothermometry, where the nanothermometers typically exhibit a monotonic temperature dependence and unique trend.

Likewise, BSA-D-MNPs exhibit the same type of profile (Figure SI1). In this case, the observed temperature sensitivity remains constant, peaking at 60 kHz.

We have also explored the thermal response of the MNPs in solutions of different viscosity and medium composition. Figure 5 shows results obtain for MNPs dispersed in media with different fractions of glycerol dispersed in DDW (0 and 66%), and in buffer with BSA dispersed at 10 mg/mL BSA solution. These media were selected to assess key aspects of intracellular [28] or protein corona formation in blood plasma [27]. The corresponding results are shown in the following figure, together with those obtained for bare MNPs for comparison.

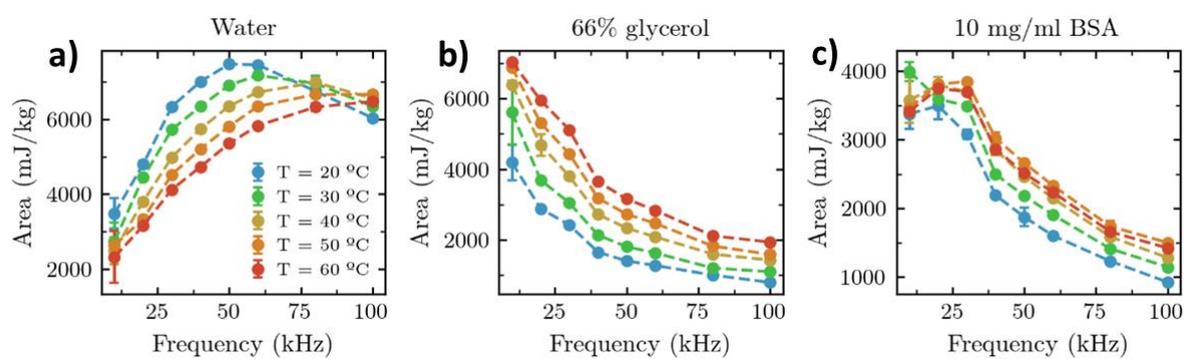

**Figure 5. Frequency dependence of the AC magnetic hysteresis-loop area of MNPs dispersed in different media.** (a) DDW 0% glycerol/0mg/ml BSA, (b) DDW 66% glycerol (v/v), (c) BSA (10 mg/mL) dispersed in PBS. Measurements were performed at different temperatures 20–60°C at 24 kA/m over the 10–100 kHz frequency range and the same magnetic element content (1 g/L). Data are shown as mean ± SD from three technical replicates (no biological replicates).

Interestingly, glycerol and BSA rich media shift the AC magnetic area profile to low frequencies while preserving the temperature sensitivity of the MNPs required for nanothermometry. The shift observed in the 10 mg/mL BSA solution arises from an increase in the effective local viscosity and hydrodynamic drag experienced by the nanoparticles, which slows Brownian relaxation and shifts the magnetic response. This effect is substantially more pronounced in glycerol, where the higher viscosity value (66% glycerol) further suppresses rotational diffusion, resulting in a greater displacement of the characteristic magnetic features. These results confirm that the thermometric response is highly sensitive to the environment, with viscosity playing a dominant role in modulating the observed behavior. As shown in Figures SI3 and SI4, linear scaling of area with temperature are preserved across the entire frequency range, except at 10 kHz. The temperature sensitivity reaches a maximum at 40 kHz for nanoparticles dispersed in 10 mg/mL BSA. In contrast, in 66% glycerol, the maximum sensitivity is observed at 20 kHz and decreases progressively at higher frequencies.

We have demonstrated that dynamical magnetization cycles of MNPs are extremely sensitive to different temperatures in the biological range from 20 to 60ºC. To validate the ability of MNPs to report temperature changes induced by itself, we applied controlled heating via near-infrared irradiation and independently verified the resulting temperature increase. As we described before, these temperature changes were induced using the same nanoparticles, enabling both heat generation and temperature readout within a single system. This approach represents an important milestone, as it allows label-free temperature modulation and sensing without the need for external reporters. In this experiment, MNPs were irradiated by near-infrared (808 nm) laser to generate localized temperature changes [24]. The purpose of this experiment is not to maximize the heating efficiency, but to demonstrate that temperature changes—regardless of their origin—can be accurately retrieved from the magnetic hysteresis area of the same nanoparticles. After NIR irradiation, the sample was transferred to the AC magnetometer and the magnetic hysteresis loop was measured at 50 kHz. The corresponding hysteresis-loop area was then converted into temperature using the empirical area–temperature calibration curve obtained in Fig. 6b, where the hysteresis area was previously measured as a function of temperature under identical magnetic field conditions. In parallel, the temperature solution was measured by inserting a thermometer to be compared with the magnetometer readout. The experimental setup and the results are shown in Figure 6.

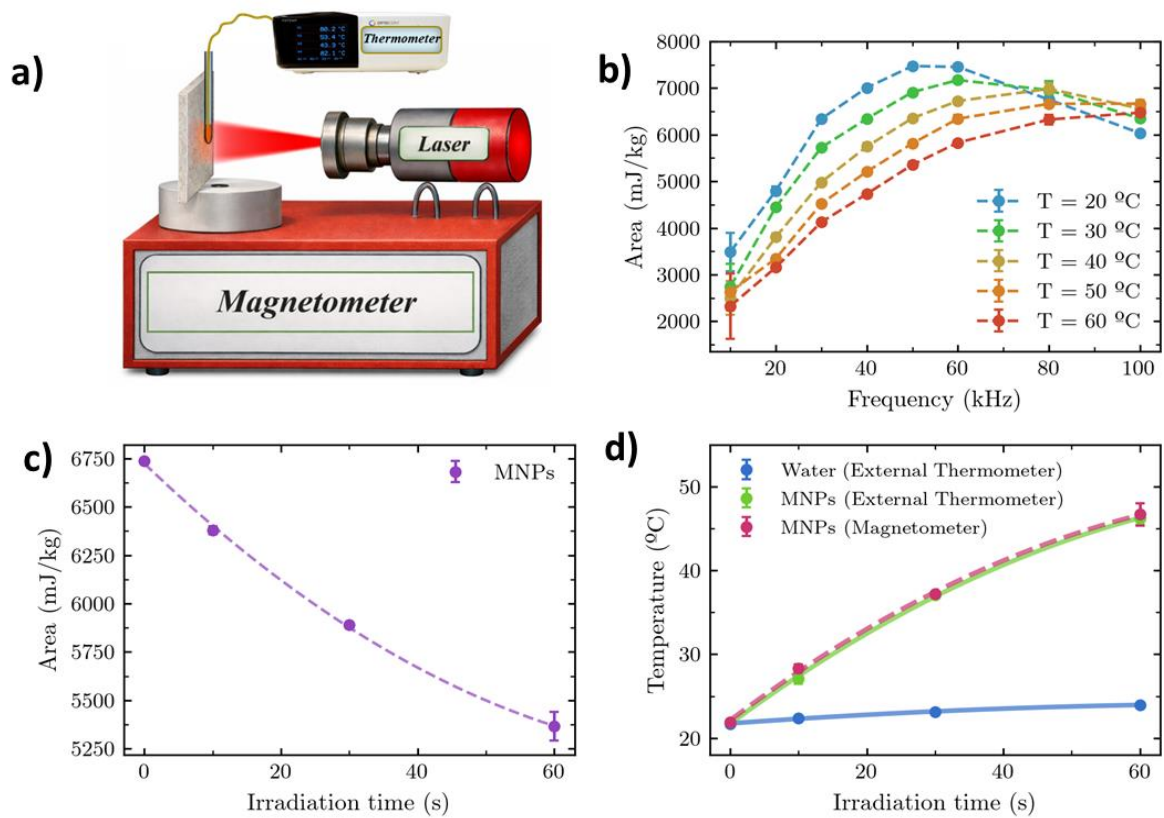

**Figure 6. Demonstration of simultaneous heat generation and temperature readout using MNPs. (a)** Schematic illustration of the experimental setup combining near-infrared (NIR, 808 nm) laser irradiation with AC magnetic hysteresis measurements. **(b)** AC Magnetic hysteresis-loop area as a function of field frequency measured at different temperatures (20–60°C), used as a calibration to extract temperature information from the magnetic response. **(c)** Evolution of the AC magnetic hysteresis area as a function of NIR irradiation time for MNPs, indicating laser-induced heating. **(d)** Comparison between the bulk solution temperature measured with an external fiber-optic probe (green line) and the temperature inferred from the AC magnetic hysteresis area of the MNPs using the calibration in (b) (black square) after NIR irradiation, demonstrating quantitative agreement between independent thermometry methods. The temperature change of pure water under identical irradiation conditions (light-blue line) is included as a control to confirm that heating originates from the MNP.

Figure 6a shows the experimental setup employed. As illustrated, a laser was used to induce temperature changes, and NIR irradiation resulted in a measurable temperature increase in the suspension containing MNPs. The solution temperature was measured with an external thermometer inserted immediately after laser irradiation. To determine the temperature using the magnetic response, the sample was subsequently placed in the magnetometer. Figure 6b shows the field frequency (50 kHz) used to measure the magnetic area, which serves as the temperature-reporting parameter. Figure 6c illustrates the evolution of the magnetic area as a function of irradiation time. Finally, Figure 6d compares the solution temperature measured with the external thermometer and the temperature reported by the MNPs. A great agreement between the two measurements is observed. Control experiments performed in the absence of nanoparticles showed no detectable changes in temperature, confirming that temperature readout originates exclusively from the MNPs. Importantly, the presented thermometric methodology is independent of the heating modality and can be directly applied to magnetic hyperthermia or other heating schemes. These results clearly demonstrate the capability of MNPs to act as a label-free platform for both heat generation and temperature sensing.

**Conclusions**

We demonstrated that cobalt ferrite magnetic nanoparticles (MNPs) can function as a fully label-free dual nanoheater–nanothermometer platform, enabling simultaneous heat generation and temperature readout from the same material. By exploiting the temperature dependence of the AC magnetization measurements, local temperature variations can be accurately extracted over a defined frequency range with high sensitivity. This is thanks to the prevalence of Brownian relaxation that tailor the AC magnetic hysteresis loops, and consequently the AC magnetic area. Importantly, this thermometric capability is preserved after surface functionalization with biologically relevant coatings and under changes in viscosity and protein concentration, highlighting the robustness of the approach in biologically relevant environments. As a proof of concept, localized heating induced by near-infrared irradiation was successfully monitored via the nanoparticles' intrinsic magnetic response. More broadly, this work establishes a generalizable route for transforming widely used magnetic nanoheaters into self-reporting, label-free nanothermometers, without introducing additional sensing components, with strong potential for improving precision and control changes in temperature in biological environments. As an important validation, we also showed that the tedious experimental calibration of the area-temperature relationship (which depends on both the MNP and the environmental properties) can be avoided by using a theoretical platform for fast and accurate automated characterization of the hysteric area in Brownian MNPs [25]. This self-consistent automated method also allows for the prior estimation of the particle size and environment viscosity from experimental data at fixed temperature. Finally, the present work considers thermal equilibrium, but the protocol can be generalized to out-of-equilibrium, to measure time-dependent temperature variations between the heated nanoparticle and the environment, opening a door to puzzling submicron-scale heat transfer effects.

**Experimental section**

**ES-1 Magnetic nanoparticles**

1) Plain surface $Co_{0.3}Fe_{2.7}O_4$ nanoflowers (MNPs), (Product Code 124-02-501; Micromod Partikeltechnologie GmbH, Germany); 2) Dextran coated surface with carboxylic groups (D-MNPs), (Product Code 123-00-301; Micromod Partikeltechnologie GmbH); 3) BSA conjugated surface onto dextran coated with carboxylic groups (BSA-D-MNPs). Conjugation protocol described in section Supporting Information 3. 4) Magnetite nanoflowers, Synomag®-D (Product Code 104-56-701, Micromod Partikeltechnologie GmbH, Germany) coated with Dextran.

**ES-2 Nanocrystal size and morphology**

The MNP size and shape were evaluated by transmission electron microscopy (TEM) JEOL 2100 microscope operating at 200 kV (point resolution 0.18 nm) at Centro Biología Molecular Severo Ochoa UAM-CSIC. TEM images were examined through manual analysis of more than 150 particles randomly selected in different areas of TEM micrographs using Image-J software to obtain the mean size and size distribution.

**ES-3 Quantification of Fe + Co content in the purchased magnetic colloids.**

The Fe and Co concentration in the studied MNP magnetic suspensions was determined by inductively coupled plasma optical emission spectrometry in an ICP-OES (PerkinElmer Optima 2100 DV) at Servicio de Análisis Químico del Instituto de Ciencia de Materiales de Madrid-CSIC (Madrid, Spain).

**ES-4 Quantification of Fe content by colorimetric assay in the bioconjugated BSA-MNPs**

Aliquots were taken from the Micromod MNP stock (control) with a known Fe + Co concentration of 6.130 mg/mL, as determined by ICP-OES, and from the freshly synthesized BSA-MNPs. Eight 7 mL glass vials were prepared under a fume hood, each containing 150 µL of 65% nitric acid. Subsequently, 5 µL of the control sample and 5 µL of the BSA-MNP suspension were added to each group of four vials. The samples were left to digest overnight at room temperature under the fume hood. The next day, 2.845 mL of Type II deionized water and 2 mL of 1 M NaOH were added to each vial. The pH of every sample was then measured and adjusted to the range of 3.2–4.5, as recommended in the Spectroquant® Iron Test protocol using HCl 1 M and NaOH 1 M. From each glass vial, 0.5 mL was taken in triplicate and brought to a final volume of 5/3 mL in microcentrifuge tubes. One drop (as specified by manufacturer) of Spectroquant® Iron Test reagent was added to each microcentrifuge tube, followed by vortexing for 3 s. Subsequently, 0.3 mL from each tube was transferred, in triplicate, to a 96-well polycarbonate microplate, resulting in 36 wells for the control sample and 36 wells for the BSA-MNPs sample. Absorbance was recorded at 569 nm using a Synergy™ H4 Hybrid Multi-Mode Microplate Reader (BioTek Instruments, Winooski, VT, USA) equipped with monochromator-based fluorescence and absorbance detection modules. Measurements were performed in flat-bottom 96-well plates (Corning®). Data acquisition and analysis were carried out using Gen5™ Data Analysis Software (BioTek Instruments). Absorbance values were converted to Fe + Co concentration using a calibration curve obtained one month prior. The control sample data were used to derive a correction factor for the BSA-MNPs measurements, since the Spectroquant® reagent exhibits a slow increase in absorbance over time due to gradual chromophore degradation. Finally, the concentration data of each sample was averaged, and expressed as mean ± standard deviation. From these values, the iron concentration of the BSA-MNPs was obtained, and considering that the control sample, as determined by ICP-OES, contained 94.28% iron and 5.72% cobalt, the total Fe + Co concentration of the BSA-MNPs was subsequently calculated.

**ES-5 AC Magnetic hysteresis and temperatura measurements**

AC magnetic characterization was performed using a bench-top AC magnetometer (AC Hyster™ Series, Nanotech Solutions, Spain). The instrument generates a sinusoidal magnetic field with tunable field frequencies 10-100 kHz and field amplitudes 4-32 kA/m, enabling rapid acquisition of full AC magnetic hysteresis loops under well-controlled conditions. For all measurements, samples were placed in the calibrated detection coil, and the magnetic response was recorded as the magnetization (M) as a function of the applied AC magnetic field (H). The system provides calibrated M and H units, allowing direct extraction of quantitative magnetic parameters. The AC magnetic hysteresis area (*A*) is calculated from the measured M–H loops and used as the primary observable for thermometric analysis. The AC Hyster™ system offers high sensitivity (down to 300 nA m²) and fast acquisition times (< 5 s per loop), ensuring high signal-to-noise ratios and excellent reproducibility (>99%) across repeated measurements. All experiments were conducted in automated acquisition mode using the manufacturer's software interface, which enables precise control of field frequency, field amplitude, and data-acquisition parameters.

For temperature-dependent studies, AC hysteresis loops were recorded at selected field frequencies while externally varying the sample temperature. The temperature was then extracted from changes in the AC magnetic hysteresis area using the previously established calibration. All measurements were performed under identical instrumental conditions to ensure consistency and reproducibility. Solution temperature was measured using digital Optocon fiber optic thermometer.

Suspensions were prepared in triplicate in 0.2 mL microcentrifuge tubes at 1 g/L Fe+Co in PBS 0.1x and vortexed for 2 s. The samples were then transferred into borosilicate glass tubes and centrifuged for 5 s at 0.4g to put the suspension at the bottom of the tube. These microcentrifuge tubes were left at room temperature for 5 min at 250 rpm horizontal rotation.

**ES-6 Preparation of BSA-D-MNPs**

D-MNPs were activated by EDC/NHS, and aminoethyl maleimide; in parallel BSA was activated by Traut's reagent chemistry, both preparation in PBS 0.1x, pH 7.40. Then, they were combined, and MNPs reacted with BSA at a 500:1 protein-to-particle ratio for 45 min at R.T. The conjugates were purified and concentrated using Amicon Ultra centrifugal filters (100 kDa) by centrifugation at 2.400 × g and washed 5 times. Successful conjugation was confirmed by DLS (Z-Size and Z-Potential), Bradford Assay, and AC Magnetometry, as detailed in Supporting Information 3.

**ES-7 NTA and hydrodynamic radius measurements**

NTA was measure using Nanosight NS300 (Malvern Instruments, United Kingdom). The pristine and bioconjugated MNP suspensions at an initial concentration of 1 $g_{Fe/Fe+Co}$/L were diluted 1:5000 in PB 0.1x buffer and injected into the instrument chamber using a 1 mL syringe. Camera settings were adjusted in order to focus the objective and track the individual Brownian motion of 20-80 nanoparticles in frame. Dynamic light scattering (DLS) measurements were performed to determine the intensity, number and volume weighted hydrodynamic sizes ($D_H$) of the b-MNPs at different experimental conditions. For that purpose, we employed a Zetasizer Nano ZS90 (Malvern Instruments, United Kingdom) equipped with a 4 mW He−Ne laser operating at 633 nm as energy source, with an angle of 173° between the incident beam and the avalanche photodiode detector. DDW and PB were used as dispersion media for measuring the colloidal properties of MNPs before bioconjugation. Otherwise, the colloidal properties of b-MNPs at different MNP and analyte concentrations were studied in PB 0.1x buffer. b-MNPs were diluted to a final MNP concentration of 0.05 $g_{Fe/Fe+Co}$/L prior measurements in a commercial cuvette under an automatic scan time (three scans per measurement).

| Nanoparticles | Size (nm) |
|---|---|
| MNPs | 38.5 ± 0.1 |
| D-MNPs | 56.7 ± 0.1 |
| BSA-D-MNPs | 40.5 ± 0.3 |
| Synomag® | 30 ± 5 (characterized by the manufacturer) |

**ES-8 Protein binding efficiency by Bradford Assay**

BSA stock solution (~10 mg/mL) was prepared, and its concentration was determined using a NanoDrop™ One/OneC microvolume UV–Vis spectrophotometer (Thermo Scientific™) equipped with Acclaro™ Sample Intelligence Technology. Measurements were performed using 2 µL of sample in the BSA preset mode with the 340 nm correction option. The instrument automatically adjusts the optical pathlength (0.03–1.0 mm) and compensates for potential contaminants to ensure accurate quantification within the 190–850 nm spectral range. Each sample was measured five times, and the average value was reported. This BSA solution was then diluted to 0.2 mg/mL and remeasured to verify the concentration. Next, 200 µL of Bradford reagent (diluted 1:4 in Type II deionized water) were added to 29 wells, from which 24 will be used for the standard curve and 5 for the supernatant characterization. The used plate was a 96-well VWR Tissue Culture Plates (Art. No. 734-2782). The required volumes of 0.1x PBS and 0.2 mg/mL BSA suspension were added to the standard curve wells to achieve the desired final concentrations, prepared in triplicate, of 0.0005, 0.0010, 0.0025, 0.0050, 0.0100, 0.0200, and 0.0300 mg/mL. In the 5 wells corresponding to the supernatant sample, 50 µL of the supernatant sample were added per well. Plate was shaken for 1 minute at normal speed in a Synergy™ H4 Hybrid Multi-Mode Microplate Reader (BioTek Instruments, Winooski, VT, USA) equipped with monochromator-based fluorescence and absorbance detection modules. Then, all wells were measured at 595 nm. A second-order polynomial calibration curve was fitted to the eight BSA standard concentrations ($R^2$ = 0.99973) and used to determine, from the supernatant concentration, that 96.41% of the BSA initially added for bioconjugation to the MNPs had bound to the particles.